\documentclass[12pt]{article}
\title{}
\author{}
\date{}
\usepackage[utf8]{inputenc}
\usepackage[letterpaper,top=1in,right=1in,left=1in,bottom=1in]{geometry}
\usepackage{amsmath}
\usepackage{amsfonts}
\usepackage{amssymb}
\usepackage{graphics}
\usepackage[demo]{graphicx}
\usepackage{placeins}
\usepackage[capposition=top]{floatrow}
\usepackage{subcaption}
\usepackage{lscape}
\usepackage{pdflscape}
\usepackage{setspace}
\usepackage{standalone}
\usepackage{fancyhdr}
\usepackage{framed}
\usepackage{color}
\usepackage{tikz}
\usepackage{pgfplots}
\usepackage{longtable}
\usepackage{booktabs,caption}
\usepackage[draft]{todonotes}
\usepackage{booktabs,caption}
\usepackage[flushleft]{threeparttable}
\usepackage{bbm}

\usepackage{multirow}
\usepackage{lineno}
\usepackage{titlesec}
\usepackage{hyperref}
\usepackage{comment}
\usepackage[draft]{todonotes}
\usepackage[justification=centering]{caption}
\pgfplotsset{compat=1.15}

\usepackage{longtable}
\usepackage{array}
\usepackage{authblk}
\usepackage{pdflscape}
\usepackage{colortbl}
\usepackage{ragged2e}
\usepackage{xfp}

\usepackage{stackengine}
\usepackage{array}
\newcolumntype{L}[1]{>{\raggedright\arraybackslash}p{#1}}
\setstackEOL{\#}
\setstackgap{L}{12pt}

\usepackage[page,toc,titletoc,title]{appendix}
\usepackage{blindtext}
\usepackage{pdfpages}

\hypersetup{
	colorlinks=true, 
	linktoc=all,     
	linkcolor=blue,  
	citecolor=blue,
    urlcolor=blue
}

\usepackage[style=apa,natbib]{biblatex}
\usepackage[all]{nowidow} 
\usepackage[protrusion=true,expansion=true]{microtype} 

\DeclareMathOperator*{\E}{\mathbb{E}}

\usepackage{orcidlink}

\hypersetup{ 	
pdfsubject = {},
pdftitle = {Multilevel Metamodels},
pdfauthor = {Gilbert, Joshua B.}
}

\usepackage{setspace}

\usepackage{color}

\title{Multilevel Metamodels: Enhancing Inference, Interpretability, and Generalizability in Monte Carlo Simulation Studies}

\begin{document}

\begin{titlepage}
\author[1]{Joshua B. Gilbert\,\orcidlink{0000-0003-3496-2710}}
\author[1]{Luke W. Miratrix\, \orcidlink{0000-0002-0078-1906}}
\affil[1]{Harvard University Graduate School of Education}

\maketitle

\begin{abstract}

\noindent Metamodels, or the regression analysis of Monte Carlo simulation results, provide a powerful tool to summarize simulation findings. However, an underutilized approach is the multilevel metamodel (MLMM) that accounts for the dependent data structure that arises from fitting multiple models to the same simulated data set. In this study, we articulate the theoretical rationale for the MLMM and illustrate how it can improve the interpretability of simulation results, better account for complex simulation designs, and provide new insights into the generalizability of simulation findings. \\


\noindent Corresponding author: \href{mailto:joshua_gilbert@g.harvard.edu}{joshua\_gilbert@g.harvard.edu} \\

\footnotesize

\noindent \textbf{Author Note:} No financial support was received for the writing of this article. The authors report no conflicts of interest. The code for this article is available as an online supplemental file (\url{https://researchbox.org/2429&PEER_REVIEW_passcode=LJICWO}). The authors wish to thank Zach Himmelsbach, Ben Domingue, and seminar participants at Stanford University for their helpful comments on drafts of this paper. \\

\noindent \textbf{Author Contributions:} Conceptualization: JG, LM; Methodology: JG, LM; Software: JG; Formal Analysis: JG; Writing---original draft preparation: JG; Writing---review and editing: JG, LM

\newpage
\end{abstract}
\end{titlepage}

\doublespacing

\section{Introduction} \label{introduction}

Monte Carlo simulation is an essential method in the toolkit of the modern statistician. From its origins in the 1940s, simulation has been applied to countless theoretical and empirical problems, and provides a particularly powerful tool when analytic solutions are not tractable, such as power analysis for complex designs or comparisons of the finite sample statistical properties of novel estimators or methods. As such, simulation has been applied in a wide range of fields, from mathematical statistics to statistical physics to causal inference in the social sciences (e.g., \cite{mooney1997monte, harrison2010introduction, bonate2001brief, binder2022monte, miratrix2022using, hunter2024pump, lee2024improving, miratrix2021applied}). 

As an example, consider the task of evaluating the properties of different estimators applied to same data set, such as the use of fixed versus random effects estimators in multi-site randomized controlled trials (RCTs) \citep{miratrix2021applied}. The general approach to such a simulation study includes creating a set of simulation conditions to generate each data set (e.g., sample size, average treatment effect size, degree of treatment heterogeneity), a method for generating data given the conditions, a set of models to be applied to each data set (e.g., fixed effects, random intercepts, random slopes), and fitting each model to each simulated data set across many replications. Results such as the point estimate, its estimated standard error (SE), and its difference from the data-generating value (i.e., error) are then combined into a single data set for further analysis.

In complex simulations comparing multiple estimators across many varying simulation factors, it can be difficult to present and interpret the simulation results to understand overall trends, such as when one estimator has lower bias or greater precision than another estimator. When comparing estimator performance, typical practice is to show performance of each estimator in a table and compare the results by eye, or possibly plot performance in a figure, and again compare by eye \citep[Chapter 13]{miratrix2023monte}. Sometimes simulation results report Monte Carlo uncertainty \citep{papadopoulos2001uncertainty}, but Monte Carlo standard errors provide uncertainty for the individual performance of an estimator in a given scenario (or aggregated across scenarios), and do not provide uncertainty on any estimated \textit{differences} between estimators. Two recent reviews show that providing \textit{any} uncertainty estimates associated with simulation findings is rare, in that only 23\% \citep{siepe2024simulation} and 7\% \citep{morris2019using} of the simulation studies examined in each review reported Monte Carlo standard errors. As a result, the utility of simulation studies in improving statistical practice remains unnecessarily limited \citep{siepe2024simulation}.

\subsection{The Simulation Metamodel}

One approach to address some of the challenges listed above is to use a regression model, regressing various metrics of estimator performance onto the simulation factors. Regression modeling of simulation results is called a ``metamodel'' because the outcome in the regression is itself a result generated from another statistical model \citep{kleijnen1981regression, friedman1985validating}. The metamodel can summarize the average influence of the simulation condition predictors (e.g., sample size) on outcomes of interest (e.g., bias, power, etc.) in complex factorial simulation designs with many combinations of simulation conditions. In a metamodel with main effects only, the coefficients for the estimator in the metamodel average over the other simulation conditions and reveal if, on average, one estimator performs better or worse than another. Interaction effects between estimator and simulation condition allow differences in performance between estimators to depend on simulation features. 

Because the simulation design and data-generating process is known and fully controlled by the researcher, metamodel regression coefficients have a straightforward causal interpretation, reflecting the average change in the outcome caused by a unit change in the predictor. As such, the metamodel allows the researcher to test specific hypotheses about how simulation conditions affect the outcome of interest, such as how larger sample sizes may reduce bias, or how including random slopes in an estimator may increase the estimated SEs of an associated main effect. The metamodel therefore provides a powerful tool to summarize and interpret simulation results \citep{rosen2015structured, skrondal2000design, friedman1988metamodel}. Furthermore, metamodels need not be limited to classical regression approaches; recent reviews have highlighted the potential affordances of machine learning approaches as new frontier to improve the accuracy of metamodels in representing complex simulation conditions \citep[pp. 152-153]{do2022metamodel}.

There are, however, a few statistical concerns that need to be taken into account when using metamodels. The first is possible heteroskedasticity due to the structure of the simulation. Heteroskedasticity is routine when using metamodels for simulation results. For example, when varying sample sizes or the presence vs. absence of predictive covariates are included in the simulation conditions, some subsets of results will provide more or less noisy estimates of estimator performance than others. In this case, the usual homoskedasticity assumption of the standard OLS regression model no longer hold. For example, when examining the performance of an estimator, performance estimates for scenarios with small sample sizes will be more variable than estimates derived from large sample sizes, even if the mean outcome in both conditions is the same. This heteroskedasticity can be fairly easily addressed with weighted least squares (WLS) or heteroskedasticity-robust standard errors (SEs) that take the precision of each condition into account \citep{kleijnen1981regression}. 

There is another statistical concern that has received less attention in the metamodeling literature: the regression assumption of conditional independence. That is, when two or more models are fit to the same simulated data set, their results will be correlated because any randomness of a simulated data set drawn from the same data-generating conditions---i.e., Monte Carlo uncertainty---is shared by all models applied to that data set. Standard metamodeling approaches such OLS or WLS typically ignore the correlation that arises from this dependence. One could apply cluster-robust SEs, clustering at the dataset or simulation condition level. However, these corrections treat the dependence only as a nuisance to be corrected for rather than an object of substantive interest in its own right, and, as we discuss in this paper, therefore limits the quality of subsequent inferences in important ways.

As an alternative to cluster-robust SEs, a common and flexible approach to addressing statistical dependence is the multilevel model (MLM, also called the random effects model, hierarchical linear model, or mixed effects model) that explicitly accounts for the correlation between observations drawn from the same cluster by including random effects for cluster (here, data set or simulation condition) membership. The MLM has seen varied applications in empirical research, such as longitudinal analysis of repeated measures over time, cross-sectional nested data, educational measurement and psychometrics, and other fields \citep{raudenbush2002hierarchical, rabe2022multilevel, singer2003applied}, but appears to be quite rare as a tool to analyze simulation results in a metamodel. We propose that the multilevel metamodel (MLMM) is a powerful tool for the analysis of simulation results. We show that the MLMM better accounts for the data structure of many simulation studies and allows for interpretable estimates of the consistency and generalizability of simulation results across the tested conditions.

We are aware of only three instances of an MLMM applied to analyze the results of an simulation study.\footnote{A search of the literature for terms such as ``Multilevel Monte Carlo'' does yield results, but these refer to ``performing most simulations with low accuracy at a correspondingly low cost, with relatively few simulations being performed at high accuracy and a high cost'' \citep[p. 259]{giles2015multilevel} rather than the random effects models explored in this study.}  The first is a 1992 dissertation on random effects metamodels in microeconomics that has rarely (if ever) been cited \citep{gentner1992microeconomic}. The second is a critique of simulation practices that briefly suggests random effects models to handle dependence \citep{skrondal2000design}. The third is a more recent study in education research that compares the statistical power of four psychometric models across a range of simulation conditions and includes random intercepts for data set \citep[pp. 10-13]{gilbert2024estimating}. Furthermore, to our knowledge, no simulation tutorials or textbooks describe using the MLMM to analyze simulation results. For example, Table \ref{tab:textbooks} lists 10 Monte Carlo methods textbooks along with whether they reference Monte Carlo simulation for comparing statistical methods (versus, e.g., purely mathematical descriptions of Monte Carlo processes), regression metamodels in general, or multilevel metamodels in particular. One textbook is dedicated to metamodeling \citep{friedman2012simulation}, but makes no mention of the MLMM, another briefly mentions only OLS metamodeling \citep[Chapter 13]{miratrix2023monte}, and the remaining eight mention neither strategy.

\begin{table}
    \centering
    \begin{tabular}{p{2in}p{2in}>{\raggedright\arraybackslash}p{0.75in}p{0.75in}p{0.75in}}
         Textbook&  Author, Year&   Comparing Statistical Methods&Regression/ Metamodel&Multilevel Metamodel\\
         \hline
 The Simulation Metamodel& \cite{friedman2012simulation}&  Yes&Yes&No\\
         Designing Monte Carlo Simulations in R&  \cite{miratrix2023monte}&   Yes&Yes&No\\
         Essentials of Monte Carlo Simulation: Statistical Methods for Building Simulation Models&  \cite{thomopoulos2012essentials}&   Yes&No&No\\
         Explorations in Monte Carlo Methods&  \cite{shonkwiler2009explorations}&   Yes&No&No\\
         Monte Carlo Simulation and Resampling Methods for Social Science&  \cite{carsey2013monte}&   Yes&No&No\\
         Exploring Monte Carlo Methods&  \cite{dunn2022exploring}&   No&No&No\\
         An Introduction to Sequential Monte Carlo&  \cite{chopin2020introduction}&   Yes&No&No\\
         Introducing Monte Carlo Methods with R&  \cite{robert2010introducing}&   No&No&No\\
 Handbook of Monte Carlo Methods& \cite{kroese2013handbook}&  Yes&No&No\\
 Monte-Carlo simulation-based statistical modeling& \cite{chen2017monte}&  Yes&No&No\\
 \hline
    \end{tabular}
    \caption{Content Analysis of Monte Carlo Simulation Textbooks}
    \label{tab:textbooks}
\end{table}

\subsubsection{Illustrating Metamodeling Strategies Through a Simple Simulation  }

To illustrate the benefits of the MLMM, we use a relatively simple running example of a simulation to assess the effects of covariate adjustment in randomized controlled trials (RCTs). Regression models for causal treatment effects are mostly unbiased whether or not covariates are included \citep{lin2013agnostic}, but covariate adjustment can reduce residual variance and estimation uncertainty and increase statistical power \citep{murnane2010methods}. Imagining for pedagogical purposes that the relevant results were unknown and not analytically tractable, we will explore them with a simulation to provide a concrete illustration of the issues at play and illustrate how the MLMM can provide insights beyond what more conventional metamodeling approaches can provide.

The typical workflow for a simulation study of the above question would include the following steps:

\begin{enumerate}
    \item  Create a grid of simulation conditions, in the present case, the size of the treatment effect, the proportion of units treated, the sample size, the correlation between the covariate and outcome, and the size of the treatment by covariate interaction (which creates heteroskedasticity in models without the covariate). 
    \item For each simulation condition, repeatedly simulate a data set with three variables: the outcome, a treatment indicator, and a single covariate, based on the simulation conditions. 
    \item For each simulated data set, fit three models: ``unadjusted,'' a simple linear regression model of the outcome on the treatment, ``adjusted,'' a multiple linear regression model with main effects for treatment and the covariate, and ``interacted,'' a multiple linear regression model with an added interaction between the treatment and the covariate \citep{lin2013agnostic}. 
    \item  Repeat the process for each simulation condition and collect the results for further analysis. 
\end{enumerate}

How should the simulation results be analyzed? While the most common approach in the simulation literature is to analyze performance separately for each estimator for each simulation condition through aggregation, tabulation, or visualization, when we are most interested in the performance \emph{differences} between estimators, metamodels that allow for formal tests of these differences offer an attractive approach.

\subsection{Metamodeling Approaches} \label{mlmm}

Typically, a metamodel is a regression of some performance metric (e.g., power, bias, false positive rate, root mean squared error, or coverage rate) onto the set of simulation factors along with, when comparing multiple estimators, indicator variables for each type of estimator. For our running case study, we might first, for example, calculate the estimated power for each estimator $i$ in simulation condition $k$, obtaining  $\overline{Y}_{i.k}$ (we use the $\overline{Y}$ notation to emphasize that we are averaging over all the generated data sets $j$ in condition $k$), and then regressing the estimated power onto the simulation conditions as follows:

\begin{align}
\label{agg_ols_eq}
    \overline{Y}_{i.k} &= \beta_0 + \beta_1 \text{adjusted}_{i.k} + \beta_2 \text{interacted}_{i.k} + \mathbf{X}_k \Gamma + \varepsilon_{i.k}
\end{align}

\noindent in which $\text{adjusted}_{i.k}$ is an indicator variable for whether we are using the adjusted estimator, and $\text{interacted}_{i.k}$ is an indicator for whether we are using the interacted estimator. $\mathbf{X}_k\Gamma$ is a matrix of simulation condition-level covariates (e.g., sample size, ATE, etc.). $\beta_0$ is the mean power for the unadjusted model (when the other simulation conditions are set to their reference groups), and $\beta_1$ and $\beta_2$ represent the average change in power caused by using the adjusted or interacted estimator, respectively, on average, across all data sets and simulation conditions, and $\varepsilon_{i.k}$ represents the residual variation unexplained by the predictors in the model. Equation \ref{agg_ols_eq}  gives an overall comparison of our two estimators to baseline: the $\beta_{1}$ and $\beta_{2}$ coefficients are the average benefit to power, averaged across all simulation scenarios, from our two possible forms of adjustment.

While Equation \ref{agg_ols_eq} provides a valuable baseline to capture the average effects of the covariates of interest across all conditions, one would typically include interactions between the dummy variables for the different estimators and other data-generating parameters, such as sample size or correlation between the outcome and the covariate, in order to see how different estimators respond differently to the simulation conditions. We will return to interaction models later on, but for clarity of exposition continue with this simpler model here.

While the OLS approach is both simple to implement and to interpret, the OLS assumption that the residuals $\varepsilon_{i.k}$ are conditionally independent is guaranteed to be violated for two reasons. First, when the multiple models are fit to the same collection of data sets, any sampling variability affecting the results in the data sets are shared across models. To illustrate, consider if one of the datasets in simulation condition $k$ had an extreme outlier, causing all three estimators to deviate widely from the others. This would translate into the three estimated performance measures sharing a common perturbation. They are thus not independent. Second, the model assumes an additive effect of the simulation conditions on the performance measure. If this is not the case, then the model will be systematically over-predicting for some conditions and under-predicting for others. This again induces a correlation structure of the residual error within the simulation condition.

While these dependencies could be addressed with, for example, cluster robust SEs at the simulation condition level, we argue that leveraging this dependence through an MLMM---rather than viewing it as a nuisance---can be a promising alternative. Further, we can actually fit a three-level MMLM to the individual estimates on the individual datasets to provide further opportunities for analysis and interpretation. We first discuss the more direct extension of the classic metamodel to MLMs, and then turn to the proposed approach of working with the estimate-level data.

\subsubsection{Two-Level MLMMs for the Aggregated Results}

The concern identified above with Equation \ref{agg_ols_eq} is that there are dependencies of the estimators within a given simulation condition $k$. We can address this by extending Equation~\ref{agg_ols_eq} to include a random intercept for simulation condition, $\nu_{0k}$:
\begin{align}
\label{agg_ri_eq}
    \overline{Y}_{i.k} &= \beta_{0k} + \beta_1 \text{adjusted}_{i.k} + \beta_2 \text{interacted}_{i.k}  + \varepsilon_{i.k} \\
    \varepsilon_{i.k} &\sim N(0,\sigma^2) \\
    \beta_{0k} &= \delta_{0} + \mathbf{X}_k \Gamma + \nu_{0k} \\
    \nu_{0k} &\sim N(0, \psi^2_0).
\end{align}

\noindent Not only does this model allow for the dependency structure within simulation condition, but the quantity $\psi^2_0$ is itself of interest. In particular, this measure of variation is a combination of the shared error of the performance metric estimates (the data dependency induced by shared datasets) and the systematic variation unexplained by the simulation parameters and estimation method. Assuming the number of simulation trials is reasonable, the bulk of this quantity will be this latter value, which tells us how much systematic variation in performance we are not explaining with our model. As we will see, this can give us important clues as to what interactions we should include between our covariates. 

Equation \ref{agg_ri_eq} is quite general. We can take any performance measure of interest and regress it on the dummy variables and simulation condition variables with random effects for simulation condition, and we immediately obtain statistically valid estimates of the contrast between our estimators that account for potentially complex interactions among the simulation conditions. 

Perhaps the most obvious benefit of the MLMM is that it can increase the precision of the estimated coefficients that contrast our estimators, when compared to the OLS approach. Consider $\hat\beta_1$ derived from Equation \ref{agg_ols_eq}, the estimated improvement in $\overline{Y}_{i.k}$ of the adjusted model over the unadjusted model. Ignoring the main effects of simulation condition included as covariates for now, the OLS metamodel is equivalent to a two-sample \textit{t}-test, or a linear regression of a continuous outcome on a dichotomous predictor that provides the mean difference between the groups. Assuming homoskedasticity and equal sample sizes between the two groups, the SE of the mean difference is given by the familiar formula $\widehat{\text{SE}}(\beta_{1\text{OLS}}) = \frac{2\sigma}{\sqrt{n}}$, where $\sigma$ is equivalent to the RMSE of the regression. Assuming random intercepts only, the formula for $\widehat{\text{SE}}(\beta_{1\text{MLM}})$ is identical, as in a multi-site trial, where each simulation condition is equivalent to a ``site'' \citep[p. 201]{raudenbush2000statistical}. However $\sigma^2$ in the MLMM is the residual, within-simulation condition variance after the between-simulation condition variance $\psi^2_0$ has been subtracted out. Because the total residual variance is the same in both metamodels ($\sigma^2$ in the OLS metamodel and $\sigma^2 + \psi^2_0$ in the MLMM), to the extent that $\psi^2_0 \ge 0$, which, as a variance, it must be, the MLMM will always be equally or more efficient than the comparable OLS metamodel in the case of Level 1 predictors. That is, despite a general ``myth'' that MLMs always provide larger SEs than OLS \citep[p. 494]{huang2018multilevel}, the SEs of within-cluster variables are often \textit{over}-estimated in OLS when between-cluster variance is high. In the simulation case, the random intercepts MLMM can only improve statistical precision and power, which has implications for both reduced computation time and the design and analysis of simulation studies (see Appendix \ref{app:ses} for additional discussion of this point in reference to our results). 

In other words, we can conceive of a simulation study as a blocked experiment with each simulation condition as a block \citep{pashley2022block}. We can take advantage of this structure by including the blocks as random effects in our model. One could include fixed effects for similar gains, but that would prohibit inclusion of the simulation factors as covariates. We therefore argue the random effects specification is more appropriate for this application.

With these considerations in mind, we can do better by fitting a model to the individual estimates, backing out our performance measures by letting the model itself do the aggregating. This allows us to separate the error from the datasets from the systematic variation missed by our model. This could be a major concern if we are unable to run a large number of trials within each simulation condition, meaning our sampling error due to the data sets could be large. We discuss how to extend to the three-level MLMM next.

\subsubsection{Three-Level MLMM for the Individual Results}

Including a random effect for simulation condition solves the dependency problem, and also provides an estimate of how much of the variation in performance is unexplained by the simulation condition main effects. However, we can actually extract more information than that by recognizing that our initial ``level 1'' in the two-level MLMM is actually aggregated data across the simulated datasets.

We can therefore instead consider the following three-level MLMM, which more closely mirrors the structure of the simulation as a data-generating process:

\begin{align}
\label{mlm3_eq}
    Y_{ijk} &= \beta_{0jk} + \beta_{10k} \text{adjusted}_{ijk} + \beta_{20k}\text{interacted}_{ijk} + \varepsilon_{ijk} \\ 
    \beta_{0jk} &= \gamma_{00k} + \zeta_{jk} \\
    \gamma_{00k} &= \delta_{000} + \mathbf{X}_k \Gamma + \nu_{0k} \\
    \beta_{10k} &= \delta_{001} + \nu_{1k} \\
    \beta_{20k} &= \delta_{002} + \nu_{2k} \\
    \begin{bmatrix}
    \nu_{0k} \\
    \nu_{1k} \\
    \nu_{2k}
    \end{bmatrix} &\sim N\left(\begin{bmatrix}
        0 \\ 0 \\ 0
    \end{bmatrix}, \begin{bmatrix}
        \psi^2_{0} & \rho_{01} & \rho_{02} \\
        \rho_{10} & \psi^2_{1} & \rho_{12} \\
        \rho_{20} & \rho_{21} & \psi^2_{2}
    \end{bmatrix}\right) \\
    \zeta_{jk} &\sim N(0, \tau^2) \\
    \varepsilon_{ijk} &\sim N(0, \sigma^2).
\end{align}

\noindent In the three-level MLMM, we no longer have a performance measure $\overline{Y}_{i.k}$, but instead need to use some estimate-level quantity $Y_{ijk}$ which, if averaged across datasets, would produce our desired performance measure. For example, for power, $Y_{ijk}$ would be an indicator of rejecting a target hypothesis (1) or not (0). For mean squared error, $Y_{ijk} = (\widehat{\text{ATE}}_{ijk}-\text{ATE}_k)^2$, i.e. the squared error of the specific estimate. See Table~\ref{tab:metrics} for additional detail regarding which $Y_{ijk}$ to use for which performance metric. 

\begin{table}
    \centering
    \begin{tabular}{cc}
         Performance Metric&  $Y_{ijk}$\\
         \hline
         Power&   1 = reject null when $\text{ATE} > 0$\\
 False Positive Rate& 1 = reject null when $\text{ATE} = 0$\\
         Bias&  $\widehat{\text{ATE}} - \text{ATE}$ \\
         Mean Squared Error&  $(\widehat{\text{ATE}} - \text{ATE})^2$ \\
 Average Estimated Standard Error&$\widehat{\text{SE}}(\text{ATE})$ \\
 Coverage& 1 = 95\% CI includes the true value\\
 \hline
    \end{tabular}
    \caption{Estimate-level quantity needed to obtain various performance metrics}
    \label{tab:metrics}
\end{table}

The three-level MLMM provides a rich description of how the estimators vary across simulation conditions. First, $\zeta_{jk}$ captures random variation between data sets and within simulation conditions with variance $\tau^2$, and $\varepsilon_{ijk}$ captures residual differences of the estimators \textit{within} data sets, after the main effects $\beta_1$ and $\beta_2$ have been accounted for.

The random effects at Level 3 have been expanded, with random slopes for the two estimators we are comparing to the baseline of the unadjusted model. The set of random effects allow for an intercept shift capturing potential interactions among the simulation conditions ($\nu_{0k}$), as with the two-level model, and also allow the effects of our adjustment estimators to vary across simulation conditions ($\nu_{1k}, \nu_{2k}$). In other words, $\delta_{001}$ represents the average effect of ``adjusted'' over ``unadjusted'' across all simulation conditions, and $\beta_{10k} = \delta_{001} + \nu_{1k}$ is the effect for a given simulation condition $k$.

$\psi^2_1$ provides the variance of $\nu_{1k}$ across simulation conditions and can be used to assess how much variation in improvement there is for ``adjusted'' across simulation conditions. We can also use $\psi^2_1$ to generate a rough prediction interval for a range of $\beta_{10k}$ values across the tested simulation conditions, a substantively important quantity for the generalizability of simulation results that has no parametric analog in the metamodels of aggregated results.

\subsubsection{Contrasting the two-level and three-level MLMM}

There are several reasons to prefer the three-level MLMM over the two-level MLMM when analyzing simulation results. First, the three-level MLMM provides all the affordances of the two level model, assuming we can express our targeted performance metric in terms of an average of an estimate-level measure. Substantively, the random slope variances ($\psi^2_1, \psi^2_2$) have both an intuitive and practically important meaning and allow for explorations of how generalizable the contrast between possible estimators are across the tested conditions. The two-level MLMM cannot provide a similar metric of generalizability because analogous random slopes models are not identified because the outcome can be ``perfectly'' fit by the model \citep{muthen2000methodological}, thus giving up a powerful measure of the generalizability of the results compared to three-level models of the disaggregated data. However, if only average effects are of interest, aggregation provides a reasonable approach, and the variability in aggregated performance measures can be taken into account by using the inverse of the squared Monte Carlo SEs as weights.\footnote{Comparisons of the two- and three-level models are analogous to comparisons between aggregated and individual participant meta-analysis \citep{stewart2012statistical, riley2023two}}

Compared to the three-level MLMM for individual simulation results, the two-level MLMM for aggregated results is most applicable to performance metrics such as true standard error or standard error calibration because these quantities are typically defined at the aggregate level, and cannot be expressed as a simple average of estimate-level characteristics. In these circumstances, a two-level random intercepts MLMM such as Equation \ref{agg_ri_eq} provides a reasonable analytic framework. However, as we will demonstrate in Section \ref{rmse}, even metrics that are typically defined at the aggregate level such as the true SE and the RMSE can be evaluated in the three-level MLMM framework, at least to good approximation, and thus gain the added benefits of the three-level approach for these metrics as well.

We next demonstrate the affordances of the MLMM in practice, using our running example of how covariate adjustment in RCTs affects the properties of an average treatment effect estimate. 

\section{Methods}

In this section, we perform a simulation comparing our three estimators for the average treatment effect in a randomized controlled trial. We will compare various metamodels of these data, and discuss when one would be preferred over another. All simulations and analyses were conducted in R.

For the simulation itself, we include the following crossed simulation factors in a full factorial design: sample size, average treatment effect size, proportion of units treated, correlation between covariate and outcome (in the control group), and treatment by covariate interaction for a total of 243 simulation conditions, as summarized in Table \ref{tab:sim_conditions}. We generate data from each condition and fit three OLS regression models---unadjusted, adjusted, and interacted---to each simulated data set. We repeat the process 100 times for a total of 72,900 model results across 24,300 data sets. For each model, we collect the metrics listed in Table \ref{tab:metrics} for analysis with respect to the estimated average treatment effect $\widehat{\text{ATE}}$ for each model.

\begin{table}
    \centering
    \begin{tabular}{clc}
         Simulation Condition&  Variable Name&Data-Generating Values\\
         \hline
         Sample Size&  \texttt{n}&100, 300, 500\\
         Average Treatment Effect Size&  \texttt{b1}&0, 0.2, 0.4\\
         Proportion of Units Treated&  \texttt{prop\_treated}&0.5, 0.7, 0.9\\
         Correlation between Covariate and Outcome&  \texttt{b2}&0.3, 0.5, 0.7\\
         Treatment by Covariate Interaction&  \texttt{b3}&0, 0.25, 0.5\\
         \hline
    \end{tabular}
    \caption{Grid of simulation conditions}
    \label{tab:sim_conditions}
        \justify
    \footnotesize Notes: The covariate is drawn from $N(0,1)$ and the treatment indicator is drawn from a Bernoulli distribution. See our supplement for additional detail.
\end{table}

To analyze the results, we fit four metamodels as shown in Table \ref{tab:metamods}, which includes the R code used to fit each metamodel, using the \texttt{lm\_robust} function for the OLS metamodel with clustered standard errors at the simulation condition level and the \texttt{lme4} package for the MLMMs \citep{bates2014fitting}. For three-level models of the squared error, we use a generalized linear model with a square root link function estimated with \texttt{glmer}. Our focal predictor variable is \texttt{estimator}, a categorical variable indicating the unadjusted, adjusted, or interacted model, where unadjusted is the reference group.\footnote{Here, the use of unadjusted as the reference group is an intuitive choice, but the choice of reference group may be arbitrary in other simulation contexts. In such cases, contrast coding may be more interpretable.}

\begin{table}
    \centering

    \begin{tabular}{>{\centering\arraybackslash}p{1in}>{\raggedright\arraybackslash}p{1in}>{\centering\arraybackslash}p{1in}p{3in}}
        \toprule
         Model Number&   Predictors&Estimation& R Code\\
         \hline
 1& Main effects& three-level RS&\texttt{lmer(y \textasciitilde{} estimator + n + prop\_treated + b1 + b2 + b3 + (1|dataset) + (model|dgp))}\\
 2& Interaction effects& three-level RS&\texttt{lmer(y \textasciitilde{} estimator*(n + prop\_treated + b1 + b2 + b3) + (1|dataset) + (model|dgp))}\\
    3&   Interaction effects&cluster-robust OLS, aggregated& \texttt{lm\_robust(y \textasciitilde{} estimator*(n + prop\_treated + b1 + b2 + b3), agg, clusters = dgp)}\\
    4&   Interaction effects&two-level RI, aggregated& \texttt{lmer(y \textasciitilde{} estimator*(n + prop\_treated + b1 + b2 + b3) +  (1|dgp), agg)}\\
    \bottomrule
    \end{tabular}
    \caption{Metamodels fit to the simulation results}
    \justify
    \footnotesize Notes: RI = Random Intercepts, RS = Random Slopes. \texttt{agg} is a data set of aggregated results; \texttt{dat} is a data set of individual results. \texttt{y} is the relevant outcome (see Table \ref{tab:metrics}), \texttt{b1} is the average treatment effect size, \texttt{b2} is the correlation between covariate and outcome in the control group, and \texttt{b3} is the treatment by covariate interaction.
    \label{tab:metamods}
\end{table}

Model 1 is a three-level random slopes MLMM of the individual results (Eqn. \ref{mlm3_eq}), with only main effects for simulation condition factors. The random slope variances ($\psi^2_1$ and $\psi^2_2$) in Model 1 allow the effect of \texttt{estimator} to vary randomly across the simulation conditions. Large values of these variance components would suggest the presence of omitted interaction effects between the simulation conditions and \texttt{estimator}. As such, Model 2 includes the two-way interactions. A reduction in $\psi^2_1$ and $\psi^2_2$ to near 0 in Model 2 would suggest that the two-way interactions adequately account for the systematic variation in the effects of \texttt{estimator} across simulation conditions. To serve as a contrast, Models 3 and 4 both analyze the aggregated data, using cluster-robust OLS and a two-level random intercepts model, respectively, and include all two-way interactions between \texttt{estimator} and the simulation conditions. 

We fit these multiple specifications to illustrate the differences between the metamodeling approaches. In a typical simulation study, the analyst would only pursue one metamodeling strategy and would most likely begin with an interaction model. We return to issues of model building and selection in the Discussion.

\section{Results}

We highlight metamodel results for power, false positives, and squared error to illustrate the affordances of the MLMM approach. Analogous metamodels for bias, coverage, estimated standard errors, and standard error calibration are included in our supplement. For both power and false positives, we multiply the binary 0/1 outcome by 100 so that the coefficients can be interpreted in percentage point terms. We treat all covariates as categorical variables so we make no linearity assumptions, though the models for binary responses could easily be extended to logit or probit specifications if desired. We spend most of our exposition on the metamodels for power to clearly describe the model-building strategy and interpretation of the results, which we then replicate with the additional examples as shorter illustrations of the same approach.

\subsection{Power} \label{power}

\begin{figure}
    \centering
    \includegraphics[width=.75\linewidth]{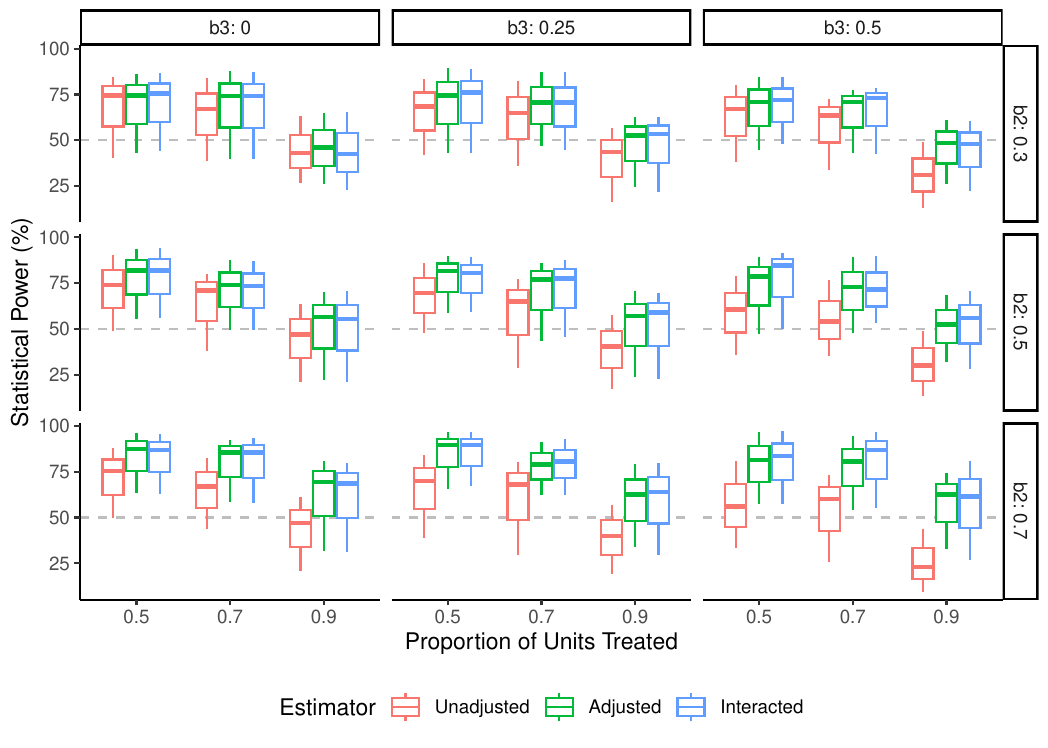}
    \caption{Distributions of Statistical Power by Simulation Condition}

    \justify \footnotesize
    The figure shows the distribution of statistical power by condition. Unadjusted represents a model with the treatment indicator only, Adjusted includes the covariate as a main effect, and Interacted includes the covariate as an interaction effect. \texttt{b3} represents the interaction between treatment and covariate and \texttt{b2} represents the correlation between covariate and outcome in the control group. Each box shows power across the varying sample sizes and the non-zero ATEs.
    \label{fig:power}
\end{figure}

To motivate the application of the MLMM to understand the statistical power results, first consider Figure \ref{fig:power}, which shows a box plot of statistical power estimates across a range of simulation conditions (aggregated by factorial condition, not results from the individual datasets). Such visualizations are common in simulation studies. We examine only the simulation conditions in which the treatment effect is positive (i.e., \texttt{b1} = 0.2, 0.4), and we clearly see that while the unadjusted model generally demonstrates lower power than the alternatives, the magnitude of the difference depends on the simulation conditions. Furthermore, there are clearly main effects of simulation conditions. For example, increasing the proportion of units treated above 50\% has deleterious effects on power across all conditions. Even in this relatively simple case, it is difficult to determine the size, consistency, and statistical significance of the differences in power by estimator based on the box plot alone. Thus, for confirmatory data analysis that strives to inform and influence statistical practice, a metamodeling approach is necessary to quantify the differences between estimators and the precision of these differences.

Table \ref{tab:power} presents the results of the metamodels for statistical power and illustrates several key insights. First, the positive coefficients for \texttt{estimator} (adjusted and interacted) in Model 1 (which excludes the interaction terms) can be interpreted as the average increase to power caused by covariate adjustment in simulation conditions similar to those explored here, of about 12.5 percentage points on average. However, the large values of $\psi^2_1$ and $\psi^2_2$ suggest that there is substantial variability in the effect of \texttt{estimator} across simulation conditions. Specifically, $\psi_1 = 10, \psi_2 = 10.8$, suggesting that the main effects, while positive, could vary by about 10 percentage points in either direction two-thirds of the time, suggesting the presence of \texttt{estimator} by simulation condition interaction effects.
\begin{table}
\caption{Metamodels for Statistical Power}
\label{tab:power}
\centering
\scalebox{.8}{
\footnotesize
\begin{tabular}{l c c c c}
\hline
 & RS & RS & OLS Agg. & RI Agg. \\
\hline
Intercept                              & $23.51 \; (1.52)^{***}$  & $29.81 \; (1.79)^{***}$  & $29.81 \; (1.79)^{***}$  & $29.81 \; (1.77)^{***}$  \\
Adjusted                               & $12.22 \; (0.83)^{***}$  & $0.91 \; (1.96)$         & $0.91 \; (1.71)$         & $0.91 \; (1.68)$         \\
Interacted                             & $12.51 \; (0.89)^{***}$  & $-0.10 \; (2.10)$        & $-0.10 \; (1.83)$        & $-0.10 \; (1.68)$        \\
N = 300                                & $26.94 \; (1.13)^{***}$  & $25.94 \; (1.39)^{***}$  & $25.94 \; (1.52)^{***}$  & $25.94 \; (1.37)^{***}$  \\
N = 500                                & $40.91 \; (1.13)^{***}$  & $41.69 \; (1.39)^{***}$  & $41.69 \; (1.39)^{***}$  & $41.69 \; (1.37)^{***}$  \\
70\% Treated                           & $-4.52 \; (1.13)^{***}$  & $-5.52 \; (1.39)^{***}$  & $-5.52 \; (1.17)^{***}$  & $-5.52 \; (1.37)^{***}$  \\
90\% Treated                           & $-25.67 \; (1.13)^{***}$ & $-27.57 \; (1.39)^{***}$ & $-27.57 \; (1.44)^{***}$ & $-27.57 \; (1.37)^{***}$ \\
ATE = 0.4                              & $34.67 \; (0.92)^{***}$  & $36.26 \; (1.13)^{***}$  & $36.26 \; (1.13)^{***}$  & $36.26 \; (1.12)^{***}$  \\
Corr. = 0.5                            & $2.74 \; (1.13)^{*}$     & $-0.26 \; (1.39)$        & $-0.26 \; (1.40)$        & $-0.26 \; (1.37)$        \\
Corr. = 0.7                            & $6.17 \; (1.13)^{***}$   & $-1.33 \; (1.39)$        & $-1.33 \; (1.36)$        & $-1.33 \; (1.37)$        \\
Int. = 0.25                            & $-1.99 \; (1.13)$        & $-4.46 \; (1.39)^{**}$   & $-4.46 \; (1.29)^{***}$  & $-4.46 \; (1.37)^{**}$   \\
Int. = 0.5                             & $-6.11 \; (1.13)^{***}$  & $-11.30 \; (1.39)^{***}$ & $-11.30 \; (1.38)^{***}$ & $-11.30 \; (1.37)^{***}$ \\
Adjusted x N = 300                     &                          & $1.30 \; (1.52)$         & $1.30 \; (1.52)$         & $1.30 \; (1.30)$         \\
Interacted x N = 300                   &                          & $2.52 \; (1.63)$         & $2.52 \; (1.62)$         & $2.52 \; (1.30)$         \\
Adjusted x N = 500                     &                          & $-2.02 \; (1.52)$        & $-2.02 \; (1.55)$        & $-2.02 \; (1.30)$        \\
Interacted x N = 500                   &                          & $-0.89 \; (1.63)$        & $-0.89 \; (1.67)$        & $-0.89 \; (1.30)$        \\
Adjusted x 70\% Treated                &                          & $1.94 \; (1.52)$         & $1.94 \; (1.55)$         & $1.94 \; (1.30)$         \\
Interacted x 70\% Treated              &                          & $1.85 \; (1.63)$         & $1.85 \; (1.66)$         & $1.85 \; (1.30)$         \\
Adjusted x 90\% Treated                &                          & $4.33 \; (1.52)^{**}$    & $4.33 \; (1.44)^{**}$    & $4.33 \; (1.30)^{***}$   \\
Interacted x 90\% Treated              &                          & $2.83 \; (1.63)$         & $2.83 \; (1.56)$         & $2.83 \; (1.30)^{*}$     \\
Adjusted x ATE = 0.4                   &                          & $-2.89 \; (1.24)^{*}$    & $-2.89 \; (1.22)^{*}$    & $-2.89 \; (1.07)^{**}$   \\
Interacted x ATE = 0.4                 &                          & $-3.14 \; (1.33)^{*}$    & $-3.14 \; (1.32)^{*}$    & $-3.14 \; (1.07)^{**}$   \\
Adjusted x Corr. = 0.5                 &                          & $5.24 \; (1.52)^{***}$   & $5.24 \; (1.09)^{***}$   & $5.24 \; (1.30)^{***}$   \\
Interacted x Corr. = 0.5               &                          & $6.17 \; (1.63)^{***}$   & $6.17 \; (1.15)^{***}$   & $6.17 \; (1.30)^{***}$   \\
Adjusted x Corr. = 0.7                 &                          & $13.93 \; (1.52)^{***}$  & $13.93 \; (1.59)^{***}$  & $13.93 \; (1.30)^{***}$  \\
Interacted x Corr. = 0.7               &                          & $14.52 \; (1.63)^{***}$  & $14.52 \; (1.74)^{***}$  & $14.52 \; (1.30)^{***}$  \\
Adjusted x Int. = 0.25                 &                          & $4.48 \; (1.52)^{**}$    & $4.48 \; (1.35)^{**}$    & $4.48 \; (1.30)^{***}$   \\
Interacted x Int. = 0.25               &                          & $4.89 \; (1.63)^{**}$    & $4.89 \; (1.42)^{***}$   & $4.89 \; (1.30)^{***}$   \\
Adjusted x Int. = 0.5                  &                          & $9.07 \; (1.52)^{***}$   & $9.07 \; (1.51)^{***}$   & $9.07 \; (1.30)^{***}$   \\
Interacted x Int. = 0.5                &                          & $10.63 \; (1.63)^{***}$  & $10.63 \; (1.63)^{***}$  & $10.63 \; (1.30)^{***}$  \\
\hline
AIC                                    & $477540.09$              & $477422.62$              & $$                       & $3129.86$                \\
BIC                                    & $477715.91$              & $477756.69$              & $$                       & $3263.82$                \\
Log Likelihood                         & $-238750.04$             & $-238673.31$             & $$                       & $-1532.93$               \\
Num. obs.                              & $48600$                  & $48600$                  & $486$                    & $486$                    \\
Num. groups: id                        & $16200$                  & $16200$                  & $$                       & $$                       \\
Num. groups: dgp                       & $162$                    & $162$                    & $$                       & $162$                    \\
Var: id (Intercept)                    & $970.46$                 & $970.48$                 & $$                       & $$                       \\
Var: dgp (Intercept)                   & $51.35$                  & $36.48$                  & $$                       & $27.89$                  \\
Var: dgp modelAdjusted                & $101.01$                 & $50.58$                  & $$                       & $$                       \\
Var: dgp modelInteracted               & $116.34$                 & $59.84$                  & $$                       & $$                       \\
Cov: dgp (Intercept) modelAdjusted     & $-54.04$                 & $-26.66$                 & $$                       & $$                       \\
Cov: dgp (Intercept) modelInteracted   & $-57.96$                 & $-28.97$                 & $$                       & $$                       \\
Cov: dgp modelAdjusted modelInteracted & $108.40$                 & $55.02$                  & $$                       & $$                       \\
Var: Residual                          & $591.57$                 & $591.43$                 & $$                       & $22.99$                  \\
R$^2$                                  & $$                       & $$                       & $0.94$                   & $$                       \\
Adj. R$^2$                             & $$                       & $$                       & $0.94$                   & $$                       \\
RMSE                                   & $$                       & $$                       & $7.13$                   & $$                       \\
N Clusters                             & $$                       & $$                       & $162$                    & $$                       \\
\hline

\end{tabular}


}
\begin{tablenotes}
\footnotesize
\item Notes: Standard errors in parentheses. OLS = Ordinary Least Squares, RI = Random Intercepts, RS = Random Slopes. $\psi^2_1$ and $\psi^2_2$ are indicated by ``Var: dgp\_modelAdjusted'' and ``Var: dgp\_modelInteracted,'' respectively. $^{***}p<0.001$; $^{**}p<0.01$; $^{*}p<0.05$
\end{tablenotes}
\end{table}

\begin{figure}
    \centering
    \includegraphics[width=.75\linewidth]{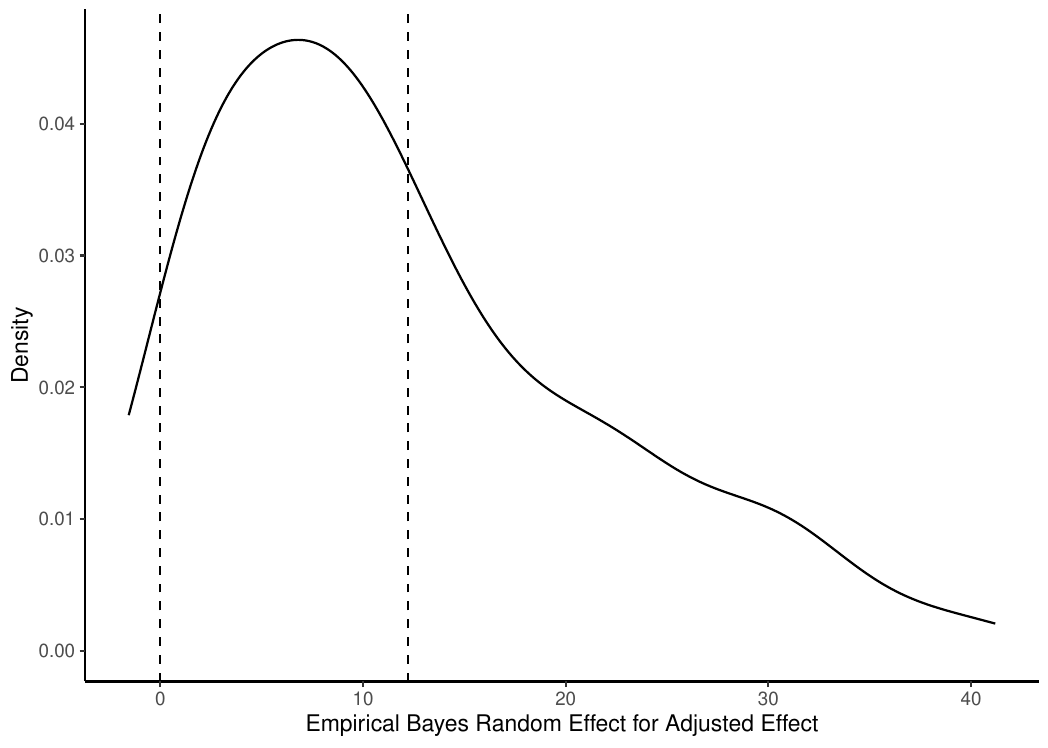}
    \caption{Distribution of Empirical Bayes Estimates for the Effect of ``Adjusted'' on Statistical Power in Model 1}

    \justify \footnotesize
        The figure shows the distribution of empirical Bayes estimates of the random slopes of ``adjusted'' across the simulated conditions from Model 1.
    \label{fig:blups}
\end{figure}

Figure \ref{fig:blups} shows the distributions of empirical Bayes estimates of the random slopes of the effect of ``adjusted'' for Model 1, and we see substantial variability. This finding suggests that the benefits or costs to statistical power of including covariates in the regression model are likely to be somewhat variable in any individual case, though nearly always positive in our sample. We can calculate the reliability of the empirical Bayes estimates using the following formula: $\frac{\psi^2}{V(\text{EB}) + \psi^2}$, where $V(\text{EB})$ is the variance of the empirical Bayes estimate. These are averaged across conditions to calculate the average reliability, yielding a reliability of .92 in Model 1. Thus, in the present case, the high reliability implies that the descriptive point estimates and the empirical Bayes estimates are very similar. In expensive simulations where only a few iterations are run per condition, unreliability and overdispersion would serve to make descriptive results such as the box plot presented in Figure \ref{fig:power} both difficult to interpret and potentially misleading compared to an MLMM that adjusts for reliability as part of the estimation procedure.

Given this variability, Model 2 includes all two-way interactions between \texttt{estimator} and the simulation conditions. The results show that the main effects of the correlation between covariate and outcome in the control group are precise nulls, which is an intuitive result because this term now reflects the effect of the predictive covariate on power for the model that includes the treatment indicator only, which we would expect to be 0. However, interactions between \texttt{estimator} and this correlation are significant, suggesting that the predictive covariate must be included in the model to increase power, as expected. The added interaction terms in Model 2 reduce the residual variance of the random slopes by about 50\% compared to Model 1. Therefore, we could continue to probe these results and explore potential three-way interactions between various combinations of metamodel parameters. If we do not include these additional interactions as fixed effects, the residual random slope variances nevertheless account for omitted interactions and adjust our estimated standard errors accordingly.\footnote{Further exploration of potential three-way interactions suggests that this additional variability is mostly driven by conditions where power is near 100\% across all estimators.}

We can also use the random slope variances to generate a 95\% prediction interval (PI) to provide a summary of the range of effects across the simulation conditions. The PI is a better metric for generalizability than the observed standard deviation of the point estimates across conditions because of overdispersion. That is, Monte Carlo uncertainty will inflate the observed variation relative to the underlying true variation, and the PI directly addresses this issue by accounting for the imprecision in the individual estimates.

Taking the ``adjusted'' effect as an example, we use the formula 
\begin{equation}
\text{PI} = \hat\beta_{10k} \pm 1.96 \sqrt{\hat{\psi^2_1} + \widehat{\text{Var}}(\beta_{10k})}
\end{equation}
\citep[p. 130]{borenstein2009introduction} to calculate the width of the 95\% PIs for $\beta_{10k}$ for Models 1 and Model 2. In Model 1, we see a very wide PI, suggesting that while the average effect of covariate adjustment on statistical power is about 12 percentage points, the effect in any individual case could be anywhere from -7 to +32 percentage points, a range of 39 percentage points. In contrast, the PI for Model 2 is significantly narrower but still large, with range of 28 percentage points, suggesting that covariate adjustment may help or hurt power in any given individual case, an important detail that would be masked by examining the average effect only.

While the prediction interval provides a useful summary measure of the generalizability of our results across simulation conditions, it also has several drawbacks. First, the interval will be symmetrical around the mean effect, which can be unrealistic for bounded metrics such as power, as we can see in Figure \ref{fig:blups}. The interval will also be equally wide in all simulation conditions due to the assumption of homoskedasticity, which is also unlikely to be tenable. As a middle ground between the over-dispersed raw SD of the point estimates and the potentially restrictive parametric assumptions of prediction interval, we use the inner 95\% range of the empirical Bayes estimates of the random slopes (subset to various combinations of our simulation condition variables, such as all cases where $N = 500$, or all cases where the proportion of treated units is 70\%) to show the variation of effects. While empirical Bayes estimates are over-shrunk relative to the truth, shrinkage is low when reliability of the estimates is high, which they are in the present case. Furthermore, the empirical Bayes estimates can potentially capture any asymmetry or heteroskedasticity in the results.

\begin{figure}
    \centering
    \includegraphics[width=.75\linewidth]{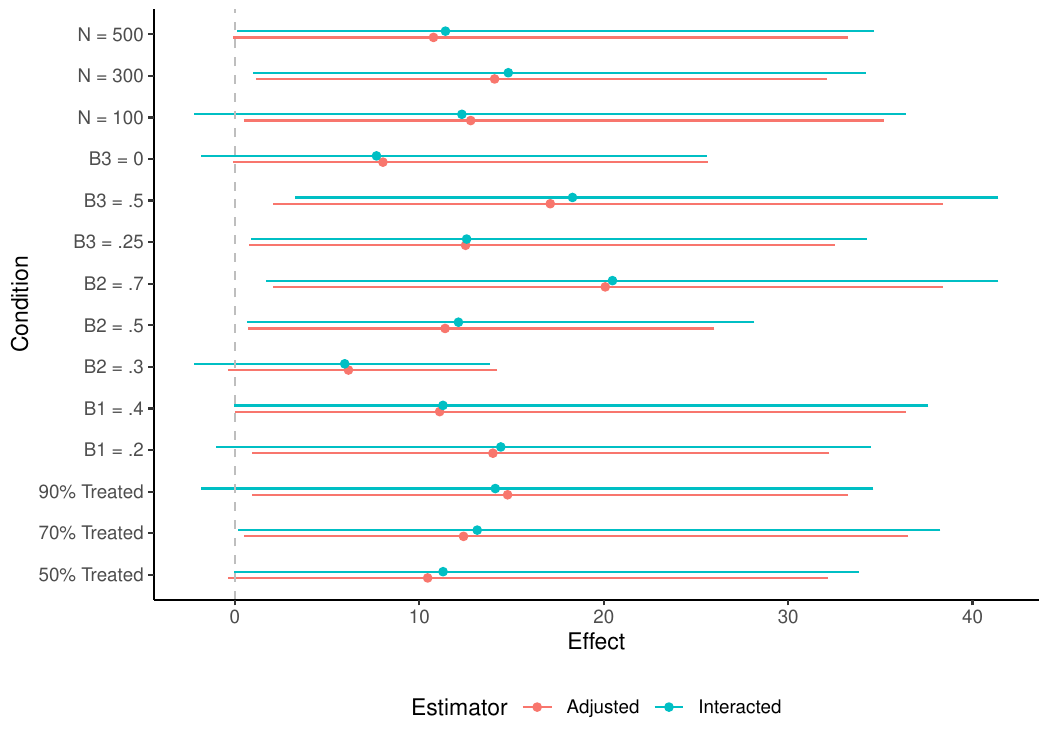}
    \caption{Conjoint Plot of Metamodel Results for Statistical Power}

    \justify \footnotesize
        The figure shows point estimates and 95\% empirical Bayes ranges for the effects of \texttt{estimator} on statistical power compared to an unadjusted model without covariates under each simulation condition. The estimates are derived from a three-level random slopes MLMM.
    \label{fig:conjoint_power}
\end{figure}

The results of the MLMM can be succinctly displayed in a conjoint plot to supplement the table, as in Figure \ref{fig:conjoint_power}. The x-axis shows the estimated effect of each estimator on power compared to the unadjusted model, derived from Model 2. The points show the average effect of covariate adjustment in each condition. The horizontal line shows the middle 95\% of empirical Bayes estimates. The y-axis shows each of the simulation condition main effects. Thus, we can see the how the effect of covariate adjustment varies according to the simulation condition.

Figure \ref{fig:conjoint_power} confirms the effect of covariate adjustment on power is generally positive across all simulation conditions. Furthermore, the 95\% empirical Bayes intervals show wide dispersion, suggesting that in any individual case, the effect of covariate adjustment could vary widely.

Finally, we test how sensitive our results are to heteroskedasticity caused by the different sample sizes of the simulation conditions, as described in Section \ref{introduction}. In our supplement, we refit the models in Table \ref{tab:metamods} applying $\sqrt{n}$ weights to account for heteroskedasticity by more heavily weighting the results from larger $n$ conditions, and we see essentially no difference in the results, suggesting that our main findings are not sensitive to heteroskedasticity. Alternative methods for addressing heteroskedasticity include the application of heteroskedasticity robust standard errors using the \texttt{sandwich} package \citep{zeileis2020various}, the \texttt{robustlmm} R package for robust MLMs \citep{koller2016robustlmm}, or a fully Bayesian approach with the \texttt{brms} R package \citep{burkner2017brms}, though the latter two methods require substantially greater computational power, taking hours or minutes rather than seconds to fit standard metamodels \citep[p. 5065]{gilbert2024tutorial}.

Models 3 and 4 present the results for the aggregated data. We see that the point estimates are identical to Model 2, and the SEs are quite similar, as we would expect due to the perfectly balanced design. However, the models of the aggregated results provide no analog of the random slope variances, and as such do not reveal how consistent the effects are across conditions, an important limitation when making recommendations for practice.

\subsection{Validity (Type I Error)}

\begin{figure}
    \centering
    \includegraphics[width=.75\linewidth]{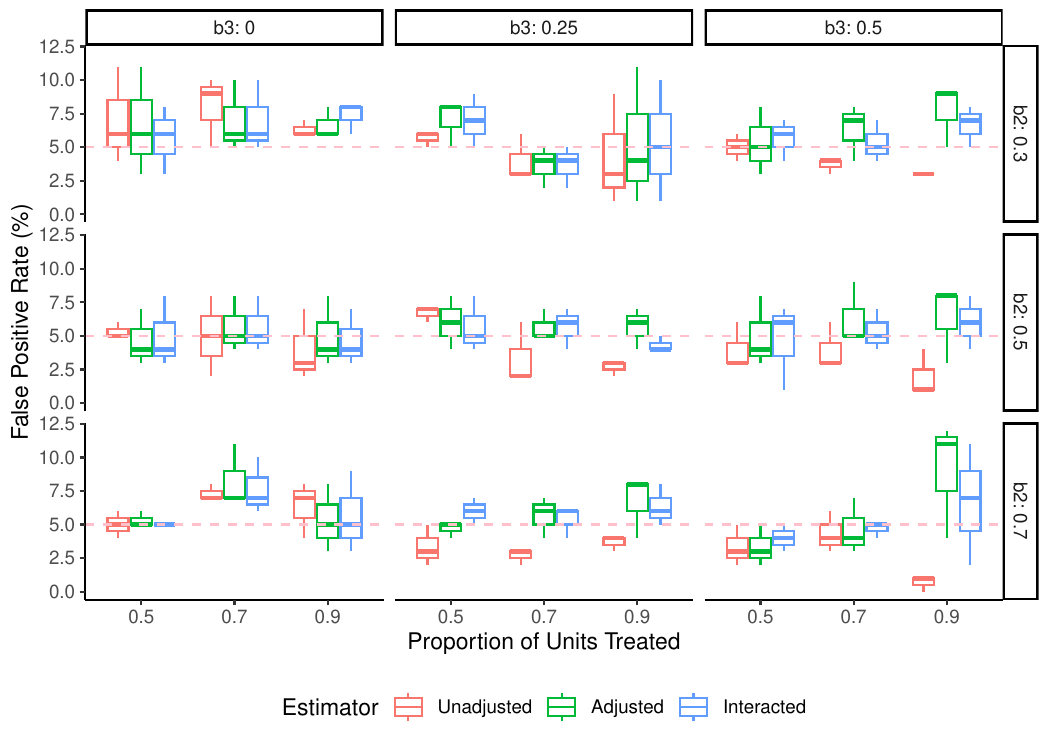}
    \caption{Distributions of False Positive Rates by Simulation Condition}
    \justify \footnotesize
        The figure shows the distribution of false positive rates by condition. Unadjusted represents a model with the treatment indicator only, Adjusted includes the covariate as a main effect, and Interacted includes the covariate as an interaction effect. \texttt{b3} represents the interaction between treatment and covariate and \texttt{b2} represents the correlation between covariate and outcome in the control group. Each box shows results averaged over the sample sizes.
    \label{fig:fp}
\end{figure}

We again begin with a descriptive graph to motivate the metamodels of false positive rates. Figure \ref{fig:fp} shows the distribution of false positive rates across a range of conditions. In contrast to Figure \ref{fig:power}, the results here are much more consistent. The false positive rates show a narrower range, and the unadjusted model without covariates has false positive rates that are too conservative when there are large interaction effects. Again, while useful as an exploratory starting point, determining the size, consistency, and statistical significance of the differences observed in the figure necessitates a metamodeling approach that accounts for imprecision and overdispersion in the descriptive results.

\begin{figure}[p]
    \centering
    \includegraphics[width=.75\linewidth]{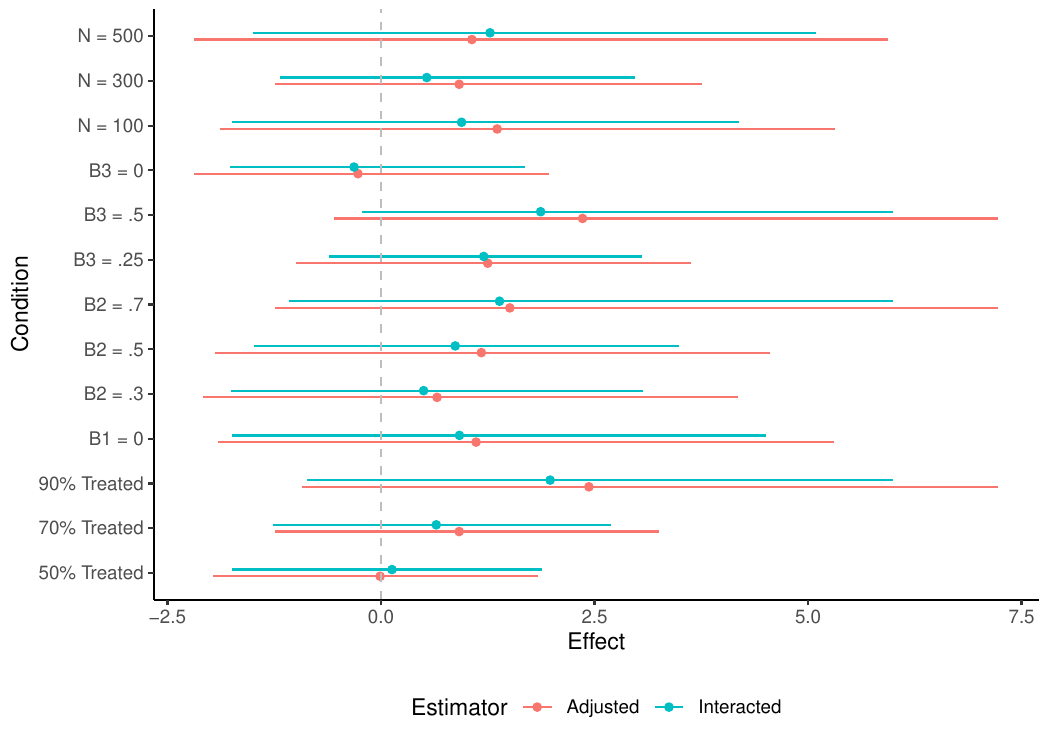}
    \caption{Conjoint Plot of Metamodel Results for False Positive Rates}

    \justify \footnotesize
        The figure shows point estimates and 95\% empirical Bayes ranges for the effects of \texttt{estimator} on the false positive rate compared to an unadjusted model without covariates under each simulation condition. The estimates are derived from a three-level random slopes MLMM.
    \label{fig:conjoint_fp}
\end{figure}

Accordingly, we replicate our metamodeling approach to power analysis for false positive rates, analyzing only the data sets in which the true treatment effect is 0. The results are shown visually in Figure \ref{fig:conjoint_fp} and serve as an illustrative contrast to the power results. Tabulations of the metamodel results are included in our supplement. We see that across the various conditions, the effects of covariate adjustment are more consistent as the mean effects vary within only a few percentage points of 0 and the 95\% empirical Bayes intervals extend only a few percentage points in either direction.

\subsection{Squared Error} \label{rmse}

We conclude with an analysis of the squared error to illustrate two additional affordances to the MLMM approach. First, even some model metrics that traditionally require aggregation can be analyzed in the three-level MLMM approach. Second, the MLMM is easily extended to non-linear contexts. 

For example, consider the Root Mean Squared Error (RMSE), which combines the effects of bias and variance into a single metric, where $r$ indexes the simulation replication and $R$ is the total number of simulation replications:

\begin{align}
    \text{RMSE} = \sqrt{\frac{1}{R}\sum_{r=1}^R{(\hat\beta_1^{(r)} - \beta_1)^2}}
\end{align}

\noindent The RMSE is traditionally analyzed in an aggregation framework because taking the mean requires averaging across a range of conditions. However, we can model the squared error directly in an MLMM approach, for example, by extending Equation \ref{mlm3_eq} to include a square root link function:

\begin{align}
    \sqrt{\E(Y_{ijk})} &= \beta_{0jk} + \beta_{10k} \text{adjusted}_{ijk} + \beta_{20k} \text{interacted}_{ijk},
\end{align}
\noindent where $Y_{ijk}$ is the squared error for simulation result $i$ in data set $j$ from simulation condition $k$. 

Figure \ref{fig:conjoint_rmse} displays the results (the full table is included in the supplement, with the coefficients transformed so they are interpretable on the RMSE scale). We see that there the average effects are negative across all conditions. However, there is wide dispersion. In some cases, the 95\% empirical Bayes range includes 0, suggesting that covariate adjustment is not guaranteed to reduce RMSE when, for example, correlations are weak and sample sizes are small. In contrast, when the correlation reaches 0.5 or 0.7, covariate adjustment is essentially guaranteed to reduce RMSE. Results are similarly consistent at larger sample sizes. Furthermore, as shown in our supplement, the estimated bias is 0 across all conditions, so these results can equivalently be interpreted as a metamodel for the true SE of each estimator, because the MSE is the sum of the squared bias and the squared SE.

\begin{figure}
    \centering
    \includegraphics[width=.75\linewidth]{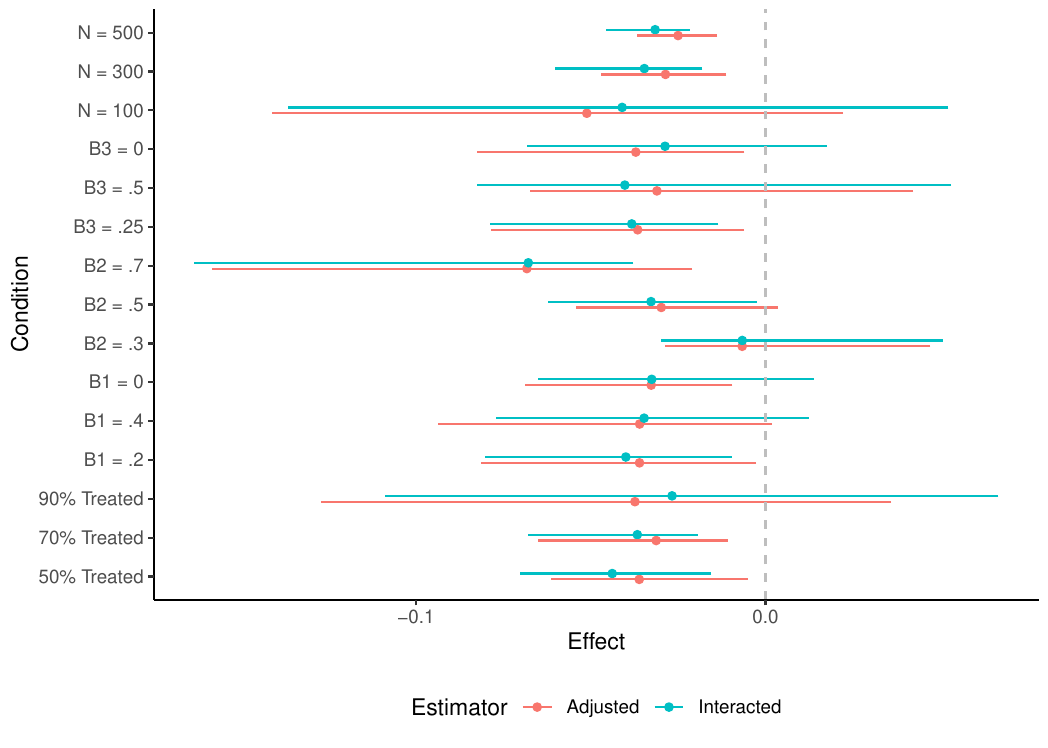}
    \caption{Conjoint Plot of Metamodel Results for Root Mean Square Error (RMSE)}

    \justify \footnotesize
        The figure shows point estimates and 95\% empirical Bayes ranges for the effects of \texttt{estimator} on the RMSE compared to an unadjusted model without covariates under each simulation condition. The estimates are derived from a three-level random slopes MLMM.
    \label{fig:conjoint_rmse}
\end{figure}

\subsection{Bias, Coverage, True Standard Error, Estimated Standard Error, and Standard Error Calibration}

Analogous metamodels in our supplement show that average bias is essentially zero across all models and simulation conditions, coverage is within a few percentage points of the nominal 95\% level, and the pattern of results observed for the relationship between the simulation conditions and the estimated standard errors and true standard errors are very similar to those of the model for the squared errors discussed earlier. Metamodels for standard error calibration show that estimated standard errors are within about 10 percentage points of the true standard errors.

\subsection{Summary and Review}

The worked examples above demonstrate how to fit and interpret the MLMM for various simulation performance metrics. The step-by-step model-building strategy outlined thus far is intended to be pedagogical by providing a side-by-side comparison of cluster-robust OLS and random intercept models of aggregated results, and random slopes models of individual results. In practice, however, the analyst would likely choose a single approach to apply to all metrics rather than compare various metamodels. We argue that the three-level random slopes MLMM with two-way interactions between the focal variable and the simulation conditions is likely to be a reasonable starting point in many applications.

\begin{figure}
    \centering
    \includegraphics[width=.75\linewidth]{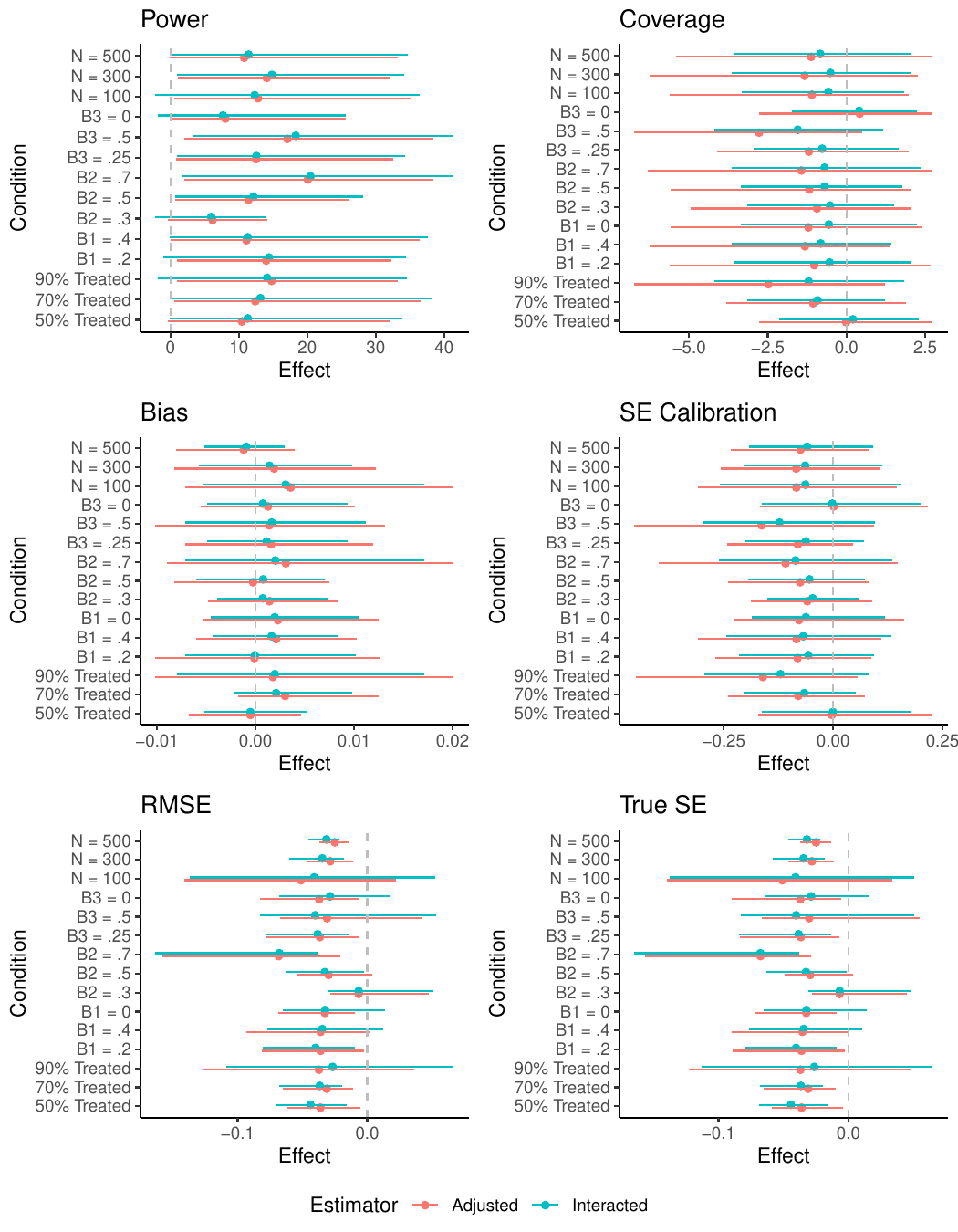}
    \caption{Conjoint Plot of Metamodel Results Across Model Metrics}
    \justify \footnotesize
      The figure shows point estimates and 95\% empirical Bayes ranges for the effects of \texttt{estimator} on various performance metrics compared to an unadjusted model without covariates under each simulation condition. The estimates are derived from a three-level random slopes MLMM.
    \label{fig:metrics}
\end{figure}

To demonstrate how an analyst might use the MLMM in practice to summarize a broad range of simulation results in practice, Figure \ref{fig:metrics} shows the conjoint plots for the average effect of covariate adjustment on bias, coverage, false positives, power, RMSE, and the true SE. The results are derived from a three-level random slopes MLMM with all two-way interactions between \texttt{estimator} and the simulation conditions. The full regression output is included in our supplement. As in the previous conjoint plots, the points show the average effect and the thin horizontal line shows the 95\% empirical Bayes range. 

Figure \ref{fig:metrics} concisely demonstrates the affordances of the random slopes MLMM as a tool for the simulation practitioner. Namely, in contrast to visualizations that are compared by eye or with descriptive statistics, the MLMM provides direct tests on the effects of the estimator on each metric in the form of \textit{differences} with a baseline model. Furthermore, the random slope model allows us to calculate a 95\% empirical Bayes range to provide an estimate of generalizability not captured by the average effect across the tested conditions conditions that are not inflated by the overdispersion of descriptive statistics and are not estimable in two-level MLMMs of aggregated results or OLS models.

How might an analyst write up the results of this simulation based on Figure \ref{fig:metrics}? We interpret the results as follows. First, we see that bias is generally small across simulation conditions: we have no evidence the average bias is not zero, and the empirical Bayes estimates extend only very narrowly in either direction. Thus, in line with statistical theory underlying randomized trials and causal inference \citep{lin2013agnostic}, covariate adjustment does not affect bias across a range of conditions. Covariate adjustment has a similarly minor effect on coverage and false positive rates, as almost all of the 95\% CIs include the mean value of the unadjusted model and the 95\% empirical Bayes ranges are relatively narrow, in the range of a few percentage points around the nominal values.

In contrast, covariate adjustment generally improves power, but benefits are strongest when the correlation between the covariate and the outcome is strong. The effects of covariate adjustment on power do not strongly depend on sample size, the proportion of units treated, or the size of the ATE. Importantly, the 95\% empirical Bayes ranges are wide, suggesting that even when correlations are strong, the effects of covariate adjustment can vary widely. For example, when sample sizes are small and the proportion of treated units is 90\%, covariate adjustment occasionally decreased power by a few percentage points.

Similarly, the RMSE and true SE show significant reductions from covariate adjustment when correlations between the covariate and the outcome are strong, and they do not increase on average in any condition. Because average bias is 0, these two figures are indistinguishable. When the correlation reaches 0.7, the 95\% empirical Bayes range does not include 0, suggesting that gains to precision are essentially guaranteed at this level of correlation in any individual case, an important interpretational nuance that would be masked if only examining the average effect.

Thus, the results of the MLMM applied to this simple simulation would suggest the following implications for analysis of RCTs in practice. Covariate adjustment can provide substantial benefits for precision (lower SEs), increased power, and lower overall error. These benefits are strongest when the correlation between the covariate and the outcome are strong and these benefits do not come at the cost of other metrics. Thus, when a predictive covariate is available, including it in the analysis either as a main effect or in an interaction with the treatment variable is generally a recommended strategy. While these are not new insights, the application of the MLMM provides in our view a succinct and interpretable method to demonstrate these findings.\footnote{To compare the ``adjusted'' and ``interacted'' estimators, we could simply change the reference group in our models and plots to allow for a direct comparison.}

Given that wide 95\% empirical Bayes ranges remain in the MLMMs for power and RMSE/SE including all two-way interaction terms, should we continue the analysis and probe three-way interactions to better understand under which fine-grained combinations of conditions covariate adjustment affects model performance? This question raises the inherent trade off between flexibility and interpretability. While significant three-way interactions undoubtedly exist in this simulation, the more flexible the model becomes, the more difficult it can be to make recommendations for practice because the results become highly sensitive to the specific conditions. One benefit of the MLMM approach is that in providing 95\% empirical Bayes ranges, simulation studies can clearly demonstrate the consistency of results across a range of settings even with higher-order interactions omitted. Thus, even if three-way interactions exist, the fact that the 95\% empirical Bayes ranges for the RMSE and the true SE do not include 0 when the correlation between covariate and outcome is 0.7 suggests that so long as this correlation is strong, gains to precision are essentially guaranteed across a wide range of simulation conditions.

\section{Discussion}

Regression metamodels of Monte Carlo simulation results have long history, but as of yet have not been widely advocated in contemporary simulation studies or textbooks. Among the many varieties of metamodels, the multilevel metamodel (MLMM) appears to be extremely rare, as we were able to identify only three extant examples in the empirical literature. In this study, we demonstrate the affordances of the metamodel in general and of the MLMM in particular as a powerful tool for the simulation practitioner.

Compared to the OLS metamodel, the MLMM better matches the data-generating process of a simulation by accounting for the dependence of results arising from models fit to the same data set or drawn from the same simulation conditions. Furthermore, random slope MLMMs allow for estimates of the generalizability of effects beyond the tested simulation conditions by providing prediction intervals or ranges of empirical Bayes estimates. Demonstrating that simulation results are consistent across a wide range of settings is essential for simulation research that strives to influence statistical practice and data analysis more broadly, because empirical applications are likely to differ in important ways from any particular set of simulation conditions \citep{skrondal2000design}.

While the MLMM provides many benefits, how does it compare to other approaches to addressing clustered or dependent data, such as aggregation, fixed effects, or cluster robust SEs? While these three alternative approaches have many merits in other applications, we argue that the MLMM is preferable in the simulation case for several reasons. First, both fixed effects and cluster robust SEs are not well equipped to handle more than two levels of hierarchy, and the default in many simulation studies will be three, with results nested within data sets nested within simulation conditions. Second, while fixed effects provide unbiased estimates of within-cluster effects in contrast to random effects models that may suffer from cluster level confounding, this is generally not an issue in simulation because the data are (or can be designed to be) perfectly balanced in that there is no between-simulation condition variation in the proportion of data sets analyzed with a given estimator (i.e., the between effect is 0) \citep[p. 597]{curran2011disaggregation}. In other words, a typical rationale for fixed effects approaches to address potential confounding does not apply here. Third, fixed effect models provide inference only for the clusters in the sample and provide no analog for the random slope variance parameter or prediction interval, thus giving up one powerful measure of generalizability. Similarly, models of results aggregated across simulation conditions cannot include random slope terms because the models are unidentified because they perfectly fit the data.

Despite these benefits, the MLMM has several limitations that bear mention: 

\begin{enumerate}
    \item As a model-based approach to simulation analysis, inferences based on the model results depend on the tenability of the model assumptions, such as the normality of the random effects (e.g., Figure \ref{fig:blups}). However, multilevel models in general are robust to violations of normality \citep{knief2021violating, schielzeth2020robustness, bell2019fixed}. 
    \item When simulations are very large, MLMMs of individual results may become impractical due to performance constraints, especially when using non-linear models. For example, we used multilevel linear probability models for the binary power and false positive responses to ease interpretation. While multilevel logit or probit models are also applicable, they are substantially more computationally intensive. 
    \item Because conclusions are based on statistical tests in the MLMM, questionable research practices such as p-hacking may become an issue. While such concerns are well understood in primary research, their application to methodological studies such as Monte Carlo simulation is less prevalent \citep{boulesteix2020replication, pawel2024pitfalls}. This issue is particularly relevant in simulation studies because the analyst can simply make the sample size larger by running additional simulations to reduce Monte Carlo SEs. In other words, ``simulations are doomed to succeed'' \citep[Chapter 21]{miratrix2023monte}. Promising solutions to this issue include preregistration of simulation studies, which may include the relevant MLMM as part of an analysis plan \citep{siepe2024simulation}.
    \item If many tests are run, the analysis may suffer from multiple comparisons issues. While some research suggests that the empirical Bayes estimates used in this study are relatively robust to multiple comparisons \citep{sales2021effect}, alternatives in the MLMM case could include replicating the simulation analysis with a different random number seed for cross validation, or the use of fully Bayesian multilevel modeling approaches \citep{burkner2017brms}.
    \item The use of the MLMM requires a model-building process. While our case study was relatively simple in its design and the hypotheses were confirmatory, other simulations may necessitate more complex model-building strategies when the goal is to discover flexible relationships between simulation design covariates and performance outcomes. As such, machine learning or model selection methods may be useful when the relevant relationships are unknown. Bayesian model averaging may also provided an attractive strategy \citep{wasserman2000bayesian}.
    \item     The arguments for the application of the MLMM as a powerful tool for simulation analysis do not encompass every simulation possibility, and thus the three-level MLMM of individual results will not always be the most appropriate tool for a given use case. For example, the presentation and worked example of this study emphasizes \textit{differences} in the performance of estimators applied to the same data set. Performance differences are not of primary importance for example when evaluating, for example, a single estimator, in which case analysis of marginal performance averaged across simulation conditions is likely sufficient. 
\end{enumerate}

In sum, the MLMM is both a powerful tool for analyzing simulation studies and easy to apply to simulation data. Simulation practitioners would be well served by exploring its potential to answer their own research questions and provide more interpretable, precise, and generalizable results to improve statistical practice in a range of applied settings.

\printbibliography

\clearpage

\appendix

\section*{Appendices}
\renewcommand{\thesubsection}{\Alph{subsection}}

\subsection{Additional Comments on Metamodel Standard Errors} \label{app:ses}

When analyzing the individual simulation results, consider the SEs provided by an OLS model with no clustering adjustments, a 3-level model with random intercepts for simulation condition, and a 3-level model with random slopes for estimator across simulation condition. While we have theoretical reasons to prefer the three-level random slopes approach, it is instructive to consider what the SEs of the OLS and random intercepts metamodels are capturing. 

We can test the calibration of the metamodel SEs by bootstrapping from our simulation results. We have three choices of which units to bootstrap: the 72,900 individual model results (a generally invalid approach because it violates independence), the 24,300 simulated data sets, or the 243 simulation conditions. We test all three of these approaches. Results show that the estimated SE provided by each metamodel essentially captures a different source of uncertainty (see \cite{abadie2020sampling}). That is, when we bootstrap the 72,900 model results, we get SEs that match those of the OLS metamodel; when we block bootstrap the 24,300 simulated datasets, we get SEs that match those of the random intercepts MLMM; when we block bootstrap the 243 simulation conditions, we get SEs that match those of the random slopes MLMM. 

These results suggest that the estimated SEs of each metamodel are capturing a different type of uncertainty in our parameter estimates \citep[p. 201]{raudenbush2000statistical}. If we are most interested in the generalizability of our results to other similar but untested simulation conditions, the random slopes MLMM provides the level of uncertainty consistent with this goal because in most simulation studies, we intend to demonstrate conclusions that we cannot prove analytically but expect or intend to hold in settings similar to the conditions tested. Thus, the random slopes MLMM is likely the most appropriate choice from both a statistical and substantive perspective. Even when there is no natural superpopulation of data-generating conditions, assessing variability across the surface defined by the tested simulation conditions nonetheless provides a useful metric for interpreting the results.

\end{document}